\input harvmac
\input epsf
\noblackbox
%%% Paragraphs
\newcount\figno
\figno=0
\def\fig#1#2#3{
\par\begingroup\parindent=0pt\leftskip=1cm\rightskip=1cm\parindent=0pt
\baselineskip=11pt
\global\advance\figno by 1
\midinsert
\epsfxsize=#3
\centerline{\epsfbox{#2}}
\vskip 12pt
\centerline{{\bf Figure \the\figno} #1}\par
\endinsert\endgroup\par}
\def\figlabel#1{\xdef#1{\the\figno}}
\def\pano{\par\noindent}

%%% special math symbols
\font\cmss=cmss10
\font\cmsss=cmss10 at 7pt

\def\th#1#2{\vartheta\bigl[{\textstyle{  #1 \atop #2}} \bigr] }
\def\rlx{\relax\leavevmode}
\def\inbar{\vrule height1.5ex width.4pt depth0pt}
\def\IC{\relax\,\hbox{$\inbar\kern-.3em{\rm C}$}}
\def\IR{\relax{\rm I\kern-.18em R}}
\def\IN{\relax{\rm I\kern-.18em N}}
\def\IP{\relax{\rm I\kern-.18em P}}
\def\ZZ{\rlx\leavevmode\ifmmode\mathchoice{\hbox{\cmss Z\kern-.4em Z}}
 {\hbox{\cmss Z\kern-.4em Z}}{\lower.9pt\hbox{\cmsss Z\kern-.36em Z}}
 {\lower1.2pt\hbox{\cmsss Z\kern-.36em Z}}\else{\cmss Z\kern-.4em Z}\fi}
\font\mengen=bbm10
\def\IOne{\hbox{\mengen 1}}
%%% misc.
\def\narrowplus{\kern -.04truein + \kern -.03truein}
\def\narrowminus{- \kern -.04truein}
\def\narrowminussub{\kern -.02truein - \kern -.01truein}

\def\o#1{\overline{#1}}

\def\th#1#2{\vartheta\bigl[{\textstyle{  #1 \atop #2}} \bigr] }

\def\pt{\partial}

%%% further macros

%%% References

\lref\witten{E. Witten, {\it BPS Bound States Of D0-D6 And D0-D8 Systems In 
A B-Field}, hep-th/0012054.}

\lref\wittenb{E. Witten, {\it Solutions Of Four-Dimensional Field Theories Via
M Theory}, Nucl.Phys. {\bf B500} (1997) 3, hep-th/9703166.}

\lref\BGKL{R. Blumenhagen, L. G\"orlich, B. K\"ors and D. L\"ust,
    {\it  Noncommutative Compactifications of Type I Strings on Tori with 
    Magnetic Background Flux},  JHEP {\bf 0010} (2000) 006,  hep-th/0007024;
    \hfil\break
    R. Blumenhagen, L. G\"orlich, B. K\"ors and D. L\"ust,  
   {\it Magnetic Flux in Toroidal Type I Compactification},
    hep-th/0010198;\hfil\break
    R. Blumenhagen, B. K\"ors and D. L\"ust,
    {\it  Type I Strings with F- and B-Flux},  hep-th/0012156.}

\lref\chen{B. Chen, H. Itoyama, T. Matsuo, K. Murakami, {\it p-p' System 
 with B-field, Branes at Angles and Noncommutative Geometry},
 Nucl.Phys. {\bf B576} (2000) 177,  hep-th/9910263.}

\lref\AADS{C. Angelantonj, I. Antoniadis, E. Dudas and A. Sagnotti, {\it
       Type-I Strings on Magnetised Orbifolds and Brane Transmutation},
       Phys.Lett. {\bf B489} (2000) 223, hep-th/0007090;  \hfil\break
     C. Angelantonj, A. Sagnotti, {\it Type-I Vacua and Brane Transmutation},
       hep-th/0010279.}

\lref\AFIQUa{G. Aldazabal, S. Franco, L. E. Ib\'a\~nez, R. Rabad\'an and 
A. M. Uranga, 
{\it D=4 Chiral String Compactifications from Intersecting Branes},
hep-th/0011073.}

\lref\AFIQUb{
G. Aldazabal, S. Franco, L. E. Ib\'a\~nez, R. Rabad\'an and A. M. Uranga, 
{\it Intersecting Brane Worlds},
hep-ph/0011132.}

\lref\miemiec{A. Karch, D. L\"ust and A. Miemiec, {\it N=1 Supersymmetric
Gauge Theories and Supersymmetric 3-Cycles}, Nucl.Phys. {\bf B553} (1999) 483,
hep-ph/9810254.}

\lref\leigh{M. Berkooz, M.R. Douglas and R.G. Leigh, 
{\it Branes Intersecting at Angles}, Nucl.Phys. {\bf B480} (1996) 265, 
  hep-th/9606139.} 

\lref\dumou{E. Dudas and J. Mourad, {\it Brane Solutions in
Strings with Broken Supersymmetry and Dilaton Tadpoles},
Phys. Lett. {\bf B486} (2000) 172, hep-th/0004165.} 

\lref\sen{A. Sen, {\it Stable Non-BPS Bound States of BPS D-branes},
JHEP {\bf 9808} (1998) 010, hep-th/9805019.}

\lref\becker{K. Becker, M. Becker and A. Strominger, {\it Fivebranes,
    Membranes and Non-Per\-tur\-bative String Theory}, 
Nucl.Phys. {\bf B456} (1995) 130, hep-th/9507158.}

\lref\douglas{G. Lawlor, {\it The Angle Criterion},
Inv. Math. {\bf 95} (1989) 437;\hfil\break
M.R. Douglas, {\it Topics in D--Geometry}, hep-th/9910170.}

\lref\park{M. Mihailescu, I.Y. Park and  T.A. Tran, {\it D-branes as Solitons
    of an N=1, D=10 Non-commutative Gauge Theory}, hep-th/0011079.}

\lref\kachru{S. Kachru and  J. McGreevy, {\it Supersymmetric Three-cycles 
and (Super)symmetry Breaking}, Phys.Rev. {\bf D61} (2000) 026001,
hep-th/9908135.}

\lref\tseyt{I. Chepelev and A.A. Tseytlin, {\it Long-distance interactions of 
branes: correspondence between supergravity and super Yang-Mills
descriptions}, Nucl.Phys. {\bf B515} (1998) 73, 
hep-th/9709087.}

%%% Title page
\Title{\vbox{
 \hbox{HU--EP--00/62}
 \hbox{hep-th/0012157}}}
{\vbox{\centerline{Bound States of D$(2p)$-D0 Systems}
\vskip 0.5cm  
\centerline{and Supersymmetric $p$-Cycles} }}
\centerline{Ralph Blumenhagen\footnote{$^1$}{{\tt 
blumenha@physik.hu-berlin.de}}, Volker Braun\footnote{$^2$}{{\tt 
volker.braun@physik.hu-berlin.de}}  and Robert Helling
\footnote{$^3$}{{\tt 
helling@ATdotDE.de}. }} 
\bigskip
\centerline{\it Humboldt-Universit\"at zu Berlin, Institut f\"ur  
Physik,}
\centerline{\it Invalidenstrasse 110, 10115 Berlin, Germany}
\smallskip
\bigskip
\centerline{\bf Abstract}
\noindent
We discuss  some issues related to D$(2p)$-D0 branes  with background
magnetic fluxes respectively, in a T-dual 
picture, Dp-Dp branes at angles. 
In particular, we describe the nature of the supersymmetric bound states 
appearing  after tachyon condensation.  
We present a  very elementary derivation of the conditions to be satisfied
by such general  supersymmetric gauge configurations, which 
are simply related by T-duality  to the conditions for 
supersymmetric $p$-cycles in $\IC^p$.

\bigskip

\Date{12/2000}
%%% text

\newsec{Introduction}

Recently, in \refs{\park,\witten} configurations of D$(2p)$-D0 branes with 
background 
fluxes were discussed.
In particular, in \witten\ the BPS bound state of a D6-D0 system in 
a non-vanishing
constant
background B-field was studied using the effective field theory on the
D0-brane. Analyzing  the preserved supercharges
for such a configuration, it was found that for a certain codimension
one sublocus the system becomes supersymmetric. On one side of
the supersymmetry locus the system is in a stable non-supersymmetric
configuration whereas it was argued that on the other side to decay
into a stable ${1\over 8}$BPS bound state. 

As was discussed in \chen, by a certain T-duality  these configurations are 
mapped to D-branes at angles \leigh, for which the supersymmetry conditions
are well known. Recently, such general configurations of D-branes
at angles were used in \refs{\BGKL,\AADS,\AFIQUa,\AFIQUb} 
to construct non-supersymmetric open 
string vacua with some appealing phenomenological properties like
chirality, supersymmetry breaking, three generations in  standard model
like gauge theories, hierarchy of Yukawa couplings and supression
of proton decay. In this context the question arose, what kind of
decay is triggered by the open string tachyons localized at the intersections
of D-branes.  The result of \witten\ indicates that via a Higgs 
mechanism in the effective theory two tachyonic intersecting D-branes 
will decay into a supersymmetric configuration. 

In this letter, we analyze the nature of these bound 
states for the D4-D0, D6-D0 and  D8-D0 systems in some more detail.
In particular, in section 2 we review the effects of a toroidal 
compactification on such D-brane configurations. 
Section 3 is devoted to an analysis of the decays of the  D$(2p)$-D0 systems.
We argue that the presence of a tachyon signals  the existence of a 
supersymmetric
configuration in the same topological sector, but with
lower energy than the D$(2p)$-D0 system. Concerning both
the number of preserved supersymmetries  and the broken
gauge group such a decay is
in agreement  with the proposed Higgs mechanism \witten\ in the effective
field theory description. Note, that the analogous  transition has been studied
in the Calabi-Yau setting in \kachru.

The equations describing such $2p$-dimensional supersymmetric gauge 
configurations
are given by a T-dual version of the conditions for supersymmetric
$p$-cycles and generalize the self-duality constraint for
BPS gauge configurations in four dimensions. 
We will derive these equations in a very elementary way
by lifting the global supersymmetry conditions, $\sum_j \Phi_j=\pi$, to 
local ones. This allows us to straightforwardly derive the 
supersymmetry conditions, which indeed turn out to be related
by T-duality to the conditions of supersymmetric $p$-cycles \miemiec\
in $\IC^p$.

\newsec{D$(2p)$-D0 branes on a torus}

In this section we first review the connection of D$(2p)$-D0 branes with 
background
fluxes and D$p$-D$p$ branes intersecting at angles.
We complexify the transversal directions of the D0-brane inside
the D$(2p)$-brane
\eqn\comp{     z_j=x_{j}+i\, y_{j} }
with $j\in\{1,\ldots,p\}$. Note that the ``real-part directions'' are given by
the D$p$-brane that was the D0-brane before the T-dualities were
performed. 
In the following we will trade  a background $B$-flux for a
background magnetic flux on the D-branes. 
A constant magnetic flux, $F$, on the D$(2p)$ brane of the form
\eqn\flux{
              F=\bigoplus_{j=1}^p 
                       \left(\matrix{0 & F_{(2j-1),2j}  \cr
                              -F_{(2j-1),2j} & 0   \cr}\right) }
is mapped via T-duality along all $p$ directions $x_{2j-1}$ into
D$p$-D$p$ branes at angles as shown for the case of D3-branes
in figure 1 
\vskip 0.5cm
\noindent
\vbox{
\hbox{\noindent\epsfysize=4.5truecm\epsfbox{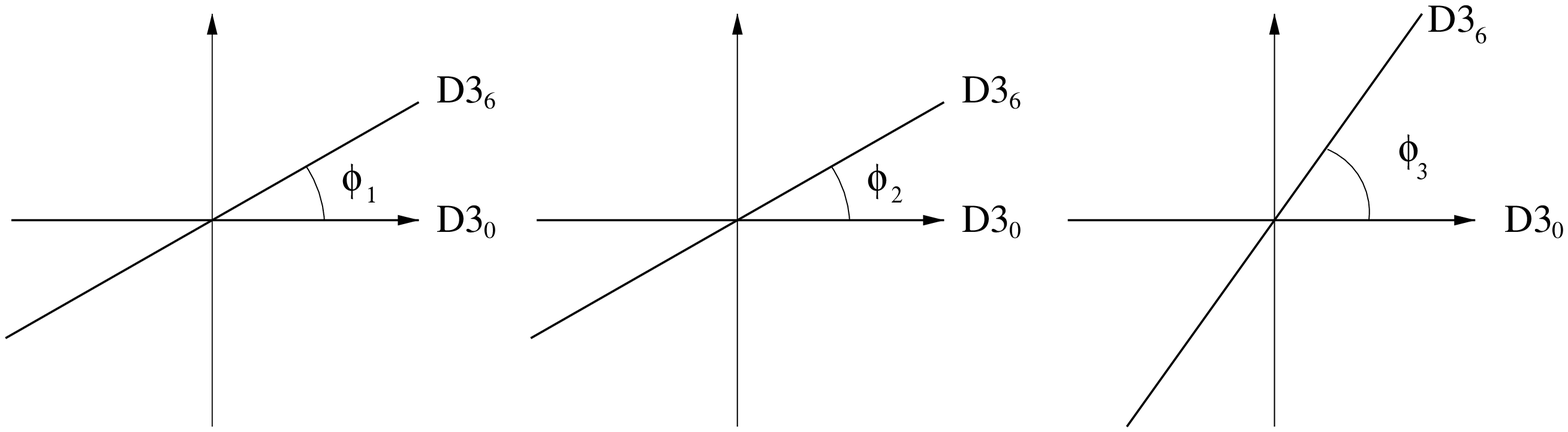}}
\noindent
\centerline{\hbox{\hskip 1.2cm Fig.1: Branes at angles }}}
\bigskip\noindent
The $p$ angles are related to the magnetic fluxes by
\eqn\rela{   F_{(2j-1),2j}=\cot \Phi_j .}
Note, that the relation between the angles $\Phi_j$ and the phases
$v_j$ in \witten\ is $\Phi_j=\pi(1/2-v_j)$. 
First compactify  the coordinates \comp\ on a torus $T^{2p}$ leading
to D$p$-D$p$ branes at angles. 
In the case of T-dual D(9-$p$)-D(9-$p$) branes filling all non-compact
directions, one has to satisfy certain non-trivial RR tadpole
cancellation conditions. In the following let us review
these conditions. 
In order to write them down, we observe that in the branes
at angles picture a D1-brane of finite length on a two-dimensional torus 
$T_{j}^2$ is described
by two wrapping numbers $(n_j,m_j)$ along the two fundamental cycles
$[{\bf 2j-1}]$ and $[{\bf 2j}]$ of the torus. 
The angle of this D1-brane with the
x-axis is given by
\eqn\winkel{    \cot\Phi_j={n_j R^{(2)}_j \over {m_j R^{(1)}_j} },}
where $R^{(1)}_j$ and $R^{(2)}_j$ denote the two radii of each $T_{j}^2$.
Under T-duality \winkel\ is mapped to the following discrete values
of the magnetic flux
\eqn\winkelb{    F_{(2j-1),2j}={n_j \over m_j R^{(1)}_j  R^{(2)}_j} ,}
so that $n_j$ can be interpreted as the magnetic charge, $c_1(F)$, of
the gauge bundle  and 
$m_j$ as the winding number of the $D2$-brane around the torus $T^2_j$.
Say, that we have $K$ different stacks of $N^{(i)}$ D-branes with
$i\in\{1,\ldots,K\}$ and that the D-branes are wrapped exactly
around one 1-cycle on each two-dimensional torus $T^2$. 
Then the total homology class of the i-th  D-brane is given by
\eqn\homo{   \Pi^{(i)}= \prod_{j=1}^p \left( n^{(i)}_j\, [{\bf 2j-1}] + 
m^{(i)}_j\,  [{\bf 2j}] \right) .}
As was first derived in the Type I case in \BGKL\ and generalized to
Type II in \AFIQUa, the RR-tadpole cancellation conditions simply mean
that the total homology class is zero
\eqn\tot{    \sum_{i=1}^K  N^{(i)} \, \Pi^{(i)} =0 .}
Expanding \tot\ yields $2^p$ non-trivial conditions for the wrapping numbers
$(n^{(i)}_j,m^{(i)}_j)$ \footnote{$^1$}{A consequence of the  condition \tot\
is that the homological class is preserved for any marginal
deformation of the D-branes. 
Strictly speaking, this was derived only for the case of D-branes filling 
all non-compact directions, however even in the general case of 
Dp-Dp branes the homological class does  not change \kachru. 
We are grateful to A. Uranga for
pointing out an error in an earlier version of this paper. }.
By T-duality the same conditions have to be
satisfied for the gauge fluxes \winkelb,  where now $(n^{(i)}_j,m^{(i)}_j)$ 
have the interpretation of  magnetic charges respectively  wrapping numbers
on $T^2_j$. 

Due to the compactification, generically two D-branes at angles
have more than one intersection point. In fact the intersection number for
two branes with wrapping numbers $(n^{(1)}_j,m^{(1)}_j)$ and
$(n^{(2)}_j,m^{(2)}_j)$ is given by
\eqn\intnr{  I_{12}=\prod_{j=1}^p \left( n^{(1)}_j\, m^{(2)}_j - m^{(1)}_j \,
   n^{(2)}_j\right) .}
Since the massless bi-fundamental chiral fermions are localized at
those intersection points, they now appear with an extra multiplicity
$I_{12}$. Due to T-duality the same extra factor must appear
for D-branes with background flux, even though in the latter case
this factor is not that obvious.

\newsec{Bound states of D$(2p)$-D0 systems}

In this section we will discuss the cases of D2-D0, D4-D0, D6-D0 and
D8-D0 
branes separately and will freely jump between the flux picture
and the more intuitive D-branes at angles picture. 
For the compact case we need some by-standing D-branes to satisfy the
tadpole condition \tot. Nevertheless, for analyzing the decay we can focus
on single D$(2p)$-D0 brane pairs. 

\subsec{D2-D0}
\noindent
As is clear from the D-branes at angles picture two such D1-branes
preserve  supersymmetry only when they are parallel, i.e.
$\Phi_1=0$. If they are anti-parallel, $\Phi_1=\pi$,  they describe a 
D1-$\o{\rm D}$1 brane pair. This is also evident from the annulus partition
function for open strings stretched between two D1-branes
\eqn\annu{\eqalign{    A_{D1,D1}= \int_0^\infty {dt\over t^5} 
{1\over 2} \, 
      \sum_{\alpha,\beta\in\{0,1/2\}} &(-1)^{2(\alpha+\beta)}\,
            e^{2 i \alpha  \Phi_1} \, 
            e^{i\pi/2} \, 
               { \th{-\beta}{\alpha}^3\,  
               \th{- \Phi_1/ \pi-\beta}{\alpha}\over
                 \eta^{9}\,  
               \th{-\Phi_1/\pi-1/2}{1/2} },}} 
which only vanishes for  $\Phi_1=0$.
If the two branes are not parallel, a tachyon develops  in 
the NS sector of open strings stretched between the two D-branes. 
Therefore, the 
two D-branes will decay into a new  configuration of D-branes
wrapping the same homological cycle but with less energy. 
In the case $\Phi_1\ne 0,\pi$ the decay product  is simply the flat D-brane 
with  wrapping numbers
$(n^{(1)}_j+n^{(2)}_j,m^{(1)}_j+m^{(1)}_j)$. As long as 
$\Phi_1\ne 0,\pi$, due to the triangle
inequality the resulting brane has smaller volume than 
the sum of the two volumes of the original D-branes at angles. 
Moreover, since after the decay one is left with only one 
flat brane, the configuration preserves maximal supersymmetry and is 
therefore ${1\over 2}$BPS. Note, that in accordance with the Higgs mechanism
proposed in \witten\ only the diagonal  $U(1)$  of the former
$U(1)\times U(1)$ gauge symmetry survives after 
condensation of the bi-fundamental tachyon (Higgs-field).

\subsec{D4-D0}
\noindent
Computing the NS ground state energy one finds that for
\eqn\bedia{\Phi_1+\Phi_2=0} 
the configuration is ${1\over 4}$BPS and
that for $\Phi_1+\Phi_2\ne0$ one gets  a tachyon. To simplify the
presentation, in \bedia\ we choose the angles in such a way that
we have positive signs everywhere.  
For $\Phi_1+\Phi_2=0$ this tachyon becomes marginal and deforms
the two intersecting D-branes with their singular intersection point 
into a smooth extended 2-cycle preserving the same amount of supersymmetry
\kachru. 
In the T-dual picture, this corresponds to deforming a singular
gauge bundle (zero size instanton) into a nonsingular gauge bundle
(thick instanton) with the same energy. 

Therefore, due to Sen's philosophy \sen\ in the tachyonic case we expect  
the system to decay  into a supersymmetric 2-cycle wrapping the 
homological cycle
\eqn\wraphom{    \Pi_3=\Pi_1+\Pi_2 .} 
Again, after the decay one
ends up with an  object which is  ${1\over 4}$BPS. 
In contrast to the D2-D0 case, here the stable object can not again
be a flat D2-brane, as this would preserve ${1\over 2}$ of the type II
supersymmetry. 

Equivalently, this can be seen from the following purely topological
argument: The self-intersection number $\Pi_3 \cdot \Pi_3 = 2 \Pi_1
\cdot \Pi_2 >0$. But a flat D2-brane could be moved off of itself by a
shift, therefore would have self-intersection number 0.

In the following we derive the supersymmetric 2-cycle condition
simply by lifting the global condition $\Phi_1+\Phi_2=0$ to a local one. 
Remember that the  general characterization
 of a supersymmetric $p$-cycle is that it is
a special Lagrangian submanifold, which means that for an embedding map
$i: (p-{\rm cycle})\longrightarrow \IC^p$ the two conditions
\eqn\lagran{\eqalign{   &i^*\, {\rm Im}\,\Omega =0 \cr
                        &i^*\,\omega =0 \cr }}
are satisfied \becker. In \lagran\ $\Omega=dz_1 \wedge \cdots \wedge dz_p$ 
denotes the holomorphic volume form on $\IC^p$ and 
$\omega = {1\over 2i}\sum dz_i \wedge d\bar{z}_i$ 
the standard K\"ahler form. 
Instead of starting with the conditions \lagran\ and derive the explicit
form of the first order partial differential equation as was done
in \miemiec, we start with the condition
\eqn\start{  \Phi_1+\Phi_2=0\Longleftrightarrow
    \cot\Phi_1+\cot\Phi_2={\pt x_1\over \pt y_1}+
 {\pt x_2\over \pt y_2}=0,}
where so far $x_{j}$ depends only on $y_j$ for the same index $j$. 
A cycle in general position can be descibed by the graph
\eqn\allg{     x_1=x_1(y_1,y_2), \quad\quad x_2=x_2(y_1,y_2) .}
To derive the generalization of \start\  for this case note that 
under T-duality the partial derivatives transform into the field 
strength of the gauge field. There we can apply the most general 
rotation such that we can T-dualize back. 

So put the partial derivatives into an antisymmetric matrix and 
apply rotations such that each $2\times2$-block remains off-diagonal:
\eqn\rot{ \left(\matrix{ 0 & -{\pt \tilde x_1\over \pt \tilde y_1} & 
                 0 & -{\pt \tilde x_1\over \pt \tilde y_2} \cr
                  {\pt \tilde{x}_1\over \pt \tilde {y}_1} & 0  & 
                {\pt \tilde{x}_2\over \pt \tilde{y}_1}  & 0 \cr 
                      0 & -{\pt \tilde{x}_2\over \pt \tilde{y}_1} &0  
              & -{\pt \tilde{x}_2\over \pt \tilde{y}_2} \cr
              {\pt \tilde x_1\over \pt \tilde y_2}   & 0 & 
          {\pt \tilde{x}_2\over \pt \tilde{y}_2} & 0 \cr }\right)
               =     R\,
             \left(\matrix{ 0 & -{\pt x_1\over \pt y_1} & 0 & 0 \cr
                           {\pt x_1\over \pt y_1} &0  & 0 & 0 \cr 
                           0 & 0 &0  & -{\pt x_2\over \pt y_2} \cr
                           0 & 0 & {\pt x_2\over \pt y_2} & 0 \cr }\right)\,
                       R^{-1} }
with
\eqn\rotb{   R=\pmatrix{\cos\theta&\sin\theta\cr
                        -\sin\theta&\cos\theta\cr}\otimes\IOne_2=
\left(\matrix{ \cos\theta & 0 & \sin\theta & 0 \cr
                            0 & \cos\theta & 0 & \sin\theta  \cr
                            -\sin\theta & 0 & \cos\theta & 0 \cr
                            0 & -\sin\theta & 0 & \cos\theta  \cr}\right).}
Inspection of the resulting matrix on the left hand side
of \rot\ yields the general conditions for a supersymmetric 2-cycle
(removing the tilde's)
\eqn\gencond{\eqalign{    {\pt x_1\over \pt y_2}&={\pt x_2\over \pt y_1} \cr
{\pt x_1\over \pt y_1}&=-{\pt x_2\over \pt y_2}. \cr}}
The first follows from the special form of the rotation matrix $R$ 
whereas the latter is \start. For the
case of D4-branes, this condition is invariant under the rotation \rotb. We
will see below that the corresponding equation for larger $p$ will not be
invariant but we will have to find its ``invariantization''. A counting of
parameters (a general antisymmetric $4\times 4$-matrix has six independant
entries, two of them have to vanish due to the ansatz \allg, there was one
independent entry in block-diagonal form, and we have two equations and one
angle) shows that these two equations are sufficient to characterize the
general rotated matrix \rot\ that obeys supersymmetry.

The equations \gencond\ are precisely the Cauchy-Riemann  differential
equations for an anti-holomorphic
map. Thus, by these elementary steps we have recovered the well known result,
that supersymmetric 2-cycles are (anti-)holomorphic curves. 
Applying  T-duality along the $x_j$-directions maps coordinates to
gauge fields  
\eqn\maps{   x_j\longrightarrow  A_{(2j-1)} .} 
As a consequence, the matrix \rot\ is mapped to
\eqn\fcond{  F=\left(\matrix{ 0 & F_{12}  & 0 & F_{14} \cr
                           -F_{12} &0  & F_{23} & 0 \cr 
                           0 & -F_{23} &0  & F_{34} \cr
                            -F_{14} & 0 & -F_{34} & 0 \cr }\right)}
and the relations \gencond\ are mapped to the anti-self-duality constraint 
\eqn\selfdual{ F=-(*F).} 
Thus, we realize that holomorphicity of a complex curve
and self-duality of a gauge field are related by T-duality.
Note, that by this T-duality we can  not obtain the most general form of the 
field strength, as some of the components are necessarily  zero. 
So far, for the 2-cycle we have not learned anything new, but the same
method can also be applied to derive the corresponding 3-cycle
and 4-cycle conditions and their T-dual versions. 

One might wonder how we could extract a local condition from the global
condition \wraphom\ for flat branes. But \lagran\ is an algebraic condition 
on the tangent space of the cycle. Thus, only first derivatives are 
involved (as we expect for BPS states) so that equations \gencond\ provide
us with an equivalent characterization of a supersymmetric 2-cycle.

\subsec{D6-D0}
\noindent
Computing the NS ground state energy one finds that for
\eqn\condic{ \Phi_1+\Phi_2+\Phi_3=\pi}
the configuration is ${1\over 8}$BPS and
that for $\Phi_1+\Phi_2+\Phi_3<\pi$ one gets  a tachyon in the NS sector. 
On the other side of the supersymmetry locus,  $\Phi_1+\Phi_2+\Phi_3>\pi$,
the system is tachyon-free and therefore stable.

This nicely complements the geometric picture that one has for the
intersection of two special Lagrangian planes in $\IC^3$ (see
\douglas): Precisely if $\Phi_1+\Phi_2+\Phi_3=\pi$ the two planes are
special Lagrangian, and only if $\Phi_1+\Phi_2+\Phi_3<\pi$ one can
deform the union of the two planes towards lower volume.

Again, for $\Phi_1+\Phi_2+\Phi_3=\pi$ the tachyonic mode becomes 
marginal and we expect that it deforms
the two intersecting D-branes into a smooth 3-cycle preserving 
the same amount of supersymmetry. In the T-dual picture this will
correspond to a smoothing of the gauge bundle.

In the non-compact case of two 3-planes in $\IC^3$ one can even explicitly  
write down such a deformation. To simplify the equations take
$\Phi_1=\Phi_2=\Phi_3={\pi\over 3}$, then (one connected component of)
\eqn\slagcycle{K_c = \left\{ (zx_1,zx_2,zx_3) \in \IC^3: 
    (x_1,x_2,x_3)\in S^2, {\rm Im}\,z^3=c \right\}}
is special Lagrangian and asymptotically approaches the two
planes. The intersection with the complex plane $(\IC,0,0)$
is depicted in figure 2 for different c.
\vskip 0.5cm
\noindent
\vbox{
\centerline{\hbox{\noindent\epsfysize=6truecm\epsfbox{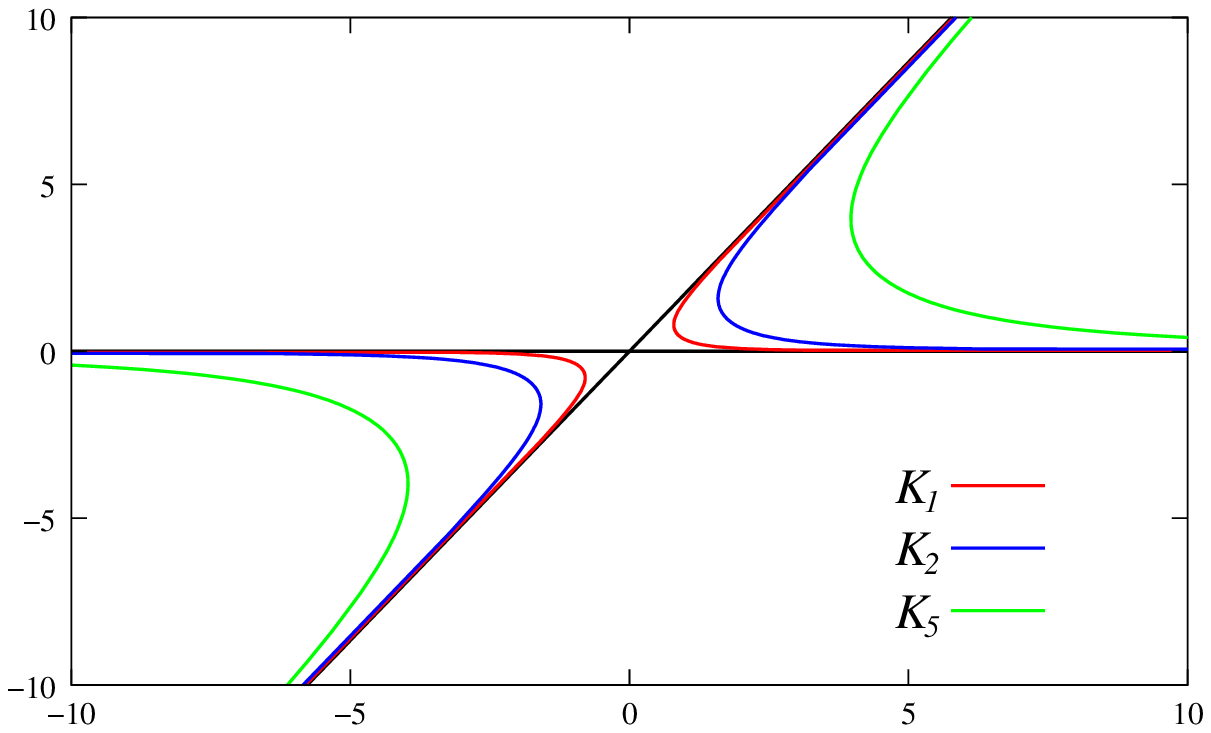}}}
\noindent
\centerline{\hbox{\hskip 1.2cm Fig.2: Deformation into a smooth 
special Lagrangian submanifold}}}
\bigskip\noindent

As in the D4-D0 case, we expect that tachyon condensation leads to
a ${1\over 8}$BPS bound state, which can be described as a necessarily
non-flat supersymmetric
3-cycle wrapping around the homological cycle $\Pi_3=\Pi_1+\Pi_2$. 
To determine the general equation satisfied by such cycles we again
require that the angle conditions are satisfied locally
\eqn\startb{
\cot\left(\Phi_1+\Phi_2+\Phi_3\right) = {\cot\Phi_1\cot\Phi_2\cot\Phi_3
-\cot\Phi_1 -\cot\Phi_2 -\cot\Phi_3\over
\cot\Phi_1\cot\Phi_2+\cot\Phi_2\cot\Phi_3+ \cot\Phi_3\cot\Phi_1-1} = -\infty }
leading to 
\eqn\fsix{   {\pt x_1\over \pt y_1}{\pt x_2\over \pt y_2}+
             {\pt x_1\over \pt y_1}{\pt x_3\over \pt y_3}+
             {\pt x_2\over \pt y_2}{\pt x_3\over \pt y_3}=1 ,}
as long as $x_j$ depends only on $y_j$.
The general 3-cycle conditions for a graph
\eqn\allg{     x_1=x_1(y_1,y_2,y_3), \quad\quad x_2=x_2(y_1,y_2,y_3) , 
               \quad\quad x_3=x_3(y_1,y_2,y_3) }
can be obtained by applying the most general $SO(3)$ rotation and
reading  off the relations for the rotated coordinates. 
\vfill\eject\noindent
We obtain
\eqn\threecyl{\eqalign{  &{\pt x_2\over \pt y_1}={\pt x_1\over \pt y_2},\quad
               {\pt x_3\over \pt y_1}={\pt x_1\over \pt y_3},\quad
               {\pt x_3\over \pt y_2}={\pt x_2\over \pt y_3}, \cr
               &{\pt x_1\over \pt y_1}{\pt x_2\over \pt y_2}+
             {\pt x_1\over \pt y_1}{\pt x_3\over \pt y_3}+
             {\pt x_2\over \pt y_2}{\pt x_3\over \pt y_3}-
             {\pt x_2\over \pt y_1}{\pt x_1\over \pt y_2}-
             {\pt x_3\over \pt y_1}{\pt x_1\over \pt y_3}-
             {\pt x_3\over \pt  y_2}{\pt x_2\over \pt y_3}=1 .}}
These four conditions agree completely with the general 
result obtained in \miemiec\ restricted to a graph \allg.
Employing T-duality we can now easily derive the generalization
of the self-duality constraint \selfdual\ to  
${1\over 8}$BPS gauge configurations. For a restricted
gauge field of the form
\eqn\fcondb{  F=\left(\matrix{ 0 & F_{12}  & 0 & F_{14} & 0 & F_{16} \cr
                          -F_{12} &0  & F_{23} & 0 & F_{25} & 0\cr 
                          0 & -F_{23} &0  & F_{34} & 0 & F_{36}\cr
                           -F_{14} & 0 & -F_{34} & 0 & F_{45} & 0 \cr 
                          0 & -F_{25} &0  & -F_{45} & 0 & F_{56}\cr
                          -F_{16} & 0 & -F_{36} & 0 & -F_{56} & 0 \cr}\right)}
the conditions for preserving  ${1\over 8}$ of the 32 supercharges are
\eqn\threeff{\eqalign{  & F_{14}=-F_{23}, \quad F_{16}=-F_{25}, \quad 
           F_{36}=-F_{45}  \cr
         & 1=F_{12}\, F_{34} +F_{12}\, F_{56} +F_{34}\, F_{56} +F_{14}\, 
           F_{23} +F_{16}\, F_{25} +F_{36}\, F_{45}. \cr }}  
As we mentioned before, the form of the constraint for the most general choice 
of the gauge field strength can not be derived by employing T-duality. 
Different ${1\over 8}$BPS gauge configurations were discussed in \tseyt.

\subsec{D8-D0}
\noindent
In this case the configuration is ${1\over 16}$BPS if
\eqn\fourangels{\Phi_1+\Phi_2+\Phi_3+\Phi_4=\pi\epsilon}
with the disjoint branches  $\epsilon=1,2$. 
For $\epsilon=2$ the system has neither
massless nor tachyonic states and will be stable. Note, that the 
usual supersymmetric D8-D0 system corresponds to
$\Phi_j=\pi/2$ for all $j$. For the other
branch $\epsilon=1$ the situation is very similar to the
D6-D0 case. For  $\Phi_1+\Phi_2+\Phi_3+\Phi_4>\pi$ the system is stable
whereas for $\Phi_1+\Phi_2+\Phi_3+\Phi_4<\pi$ it develops a tachyon.
Again, we expect that after tachyon condensation the system
will decay into a supersymmetric 4-cycle in $\IC^4$ and
the equations governing the 4-cycle 
\eqn\fourc{\eqalign{
     &{\pt x_1\over \pt y_1}{\pt x_2\over \pt y_2}{\pt x_3\over \pt y_3}+
     {\pt x_1\over \pt y_1}{\pt x_2\over \pt y_2}{\pt x_4\over \pt y_4}+
     {\pt x_1\over \pt y_1}{\pt x_3\over \pt y_3}{\pt x_4\over \pt y_4}+
     {\pt x_2\over \pt y_2}{\pt x_3\over \pt y_3}{\pt x_4\over \pt y_4}- \cr
     & {\pt x_1\over \pt y_1}-{\pt x_2\over \pt y_2}-{\pt x_3\over \pt y_3}
      -{\pt x_1\over \pt y_1}=0.}}
can be derived from  the corresponding local angle relation.
The generalization for a general graph
\eqn\allgb{\eqalign{    &x_1=x_1(y_1,y_2,y_3,y_4),
            \quad\quad x_2=x_2(y_1,y_2,y_3,y_4), \cr 
              & x_3=x_3(y_1,y_2,y_3,y_4), 
             \quad\quad x_4=x_4(y_1,y_2,y_3,y_4)} }
can be found by applying the most general $SO(4)$ rotation and extracting
the conditions
\eqn\fourcylb{\eqalign{  &{\pt x_2\over \pt y_1}={\pt x_1\over \pt y_2},\quad
               {\pt x_3\over \pt y_1}={\pt x_1\over \pt y_3},\quad
               {\pt x_4\over \pt y_1}={\pt x_1\over \pt y_4}, \
               {\pt x_3\over \pt y_2}={\pt x_2\over \pt y_3},\quad
               {\pt x_4\over \pt y_2}={\pt x_2\over \pt y_4},\quad
               {\pt x_4\over \pt y_3}={\pt x_3\over \pt y_4}, \cr
     &\cr
    &{\pt x_1\over \pt y_1}{\pt x_2\over \pt y_2}{\pt x_3\over \pt y_3}+
     {\pt x_1\over \pt y_1}{\pt x_2\over \pt y_2}{\pt x_4\over \pt y_4}+
     {\pt x_1\over \pt y_1}{\pt x_3\over \pt y_3}{\pt x_4\over \pt y_4}+
     {\pt x_2\over \pt y_2}{\pt x_3\over \pt y_3}{\pt x_4\over \pt y_4}+
     {\pt x_1\over \pt y_2}{\pt x_1\over \pt y_3}{\pt x_2\over \pt y_3}+\cr
    &{\pt x_1\over \pt y_2}{\pt x_1\over \pt y_4}{\pt x_2\over \pt y_4}+
     {\pt x_1\over \pt y_3}{\pt x_1\over \pt y_4}{\pt x_3\over \pt y_4}+
     {\pt x_2\over \pt y_3}{\pt x_3\over \pt y_4}{\pt x_4\over \pt y_3}+
     {\pt x_2\over \pt y_1}{\pt x_3\over \pt y_1}{\pt x_3\over \pt y_2}+
     {\pt x_2\over \pt y_1}{\pt x_4\over \pt y_1}{\pt x_4\over \pt y_2}+\cr
    &{\pt x_3\over \pt y_1}{\pt x_4\over \pt y_1}{\pt x_4\over \pt y_3}+
     {\pt x_3\over \pt y_2}{\pt x_4\over \pt y_3}{\pt x_3\over \pt y_4}-
     {\pt x_2\over \pt y_1}{\pt x_1\over \pt y_2}{\pt x_4\over \pt y_4}-
     {\pt x_3\over \pt y_1}{\pt x_1\over \pt y_3}{\pt x_4\over \pt y_4}-
     {\pt x_3\over \pt y_2}{\pt x_2\over \pt y_3}{\pt x_4\over \pt y_4}-\cr
    &{\pt x_2\over \pt y_1}{\pt x_1\over \pt y_2}{\pt x_3\over \pt y_3}-
     {\pt x_4\over \pt y_1}{\pt x_1\over \pt y_4}{\pt x_3\over \pt y_3}-
     {\pt x_4\over \pt y_2}{\pt x_2\over \pt y_4}{\pt x_3\over \pt y_3}-
     {\pt x_3\over \pt y_1}{\pt x_1\over \pt y_3}{\pt x_2\over \pt y_2}-
     {\pt x_4\over \pt y_1}{\pt x_1\over \pt y_4}{\pt x_2\over \pt y_2}-\cr
    &{\pt x_4\over \pt y_3}{\pt x_3\over \pt y_4}{\pt x_2\over \pt y_2}-
     {\pt x_3\over \pt y_2}{\pt x_2\over \pt y_3}{\pt x_1\over \pt y_1}-
     {\pt x_4\over \pt y_2}{\pt x_2\over \pt y_4}{\pt x_1\over \pt y_1}-
     {\pt x_4\over \pt y_3}{\pt x_3\over \pt y_4}{\pt x_1\over \pt y_1}-\cr
    & {\pt x_1\over \pt y_1}-{\pt x_2\over \pt y_2}-{\pt x_3\over \pt y_3}
      -{\pt x_1\over \pt y_1}=0.}}
In deriving these increasingly complicated expressions it proved to
be helpful to use that in the limit ${\pt x_j\over\pt y_j}\to \infty$
one gets back the 3-cycle conditions.
The T-dual conditions for a ${1\over 16}$BPS gauge configuration of the 
form
\eqn\fcondc{  F=\left(\matrix{ 
    0 & F_{12}  & 0 & F_{14} & 0 & F_{16} & 0 & F_{18} \cr
    -F_{12} &0  & F_{23} & 0 & F_{25} & 0 & F_{27} & 0 \cr 
    0 & -F_{23} &0  & F_{34} & 0 & F_{36} & 0 & F_{38}\cr
    -F_{14} & 0 & -F_{34} & 0 & F_{45} & 0 & F_{47} & 0\cr 
     0 & -F_{25} &0  & -F_{45} & 0 & F_{56} & 0 & F_{58}\cr
     -F_{16} & 0 & -F_{36} & 0 & -F_{56} & 0 & F_{67} & 0\cr
     0 & -F_{27} &0  & -F_{47} & 0 & -F_{67} & 0 & F_{78}\cr
     -F_{18} & 0 & -F_{38} & 0 & -F_{58} & 0 & -F_{78} & 0\cr
}\right)}
read
\eqn\fourff{\eqalign{  & F_{14}=-F_{23}, \quad F_{16}=-F_{25}, \quad 
           F_{18}=-F_{27},\quad
           F_{36}=-F_{45}, \quad F_{38}=-F_{47}, \quad 
           F_{58}=-F_{67},  \cr
          &\cr
         & F_{12}\, F_{34}\, F_{56} + F_{12}\, F_{34}\, F_{78} +
           F_{12}\, F_{56}\, F_{78} +F_{34}\, F_{56}\, F_{78}+ 
          F_{14}\, F_{16}\, F_{36} + \cr
         & F_{14}\, F_{18}\, F_{38} + F_{16}\, F_{18}\, F_{58} + 
           F_{36}\, F_{38}\, F_{58} - 
          F_{23}\, F_{25}\, F_{45} - F_{23}\, F_{27}\, F_{47} - \cr
         &  F_{25}\, F_{27}\, F_{67} - F_{45}\, F_{47}\, F_{67}  +
          F_{14}\, F_{23}\, F_{78} + F_{16}\, F_{25}\, F_{78} +
           F_{36}\, F_{45}\, F_{78} + \cr
         & F_{14}\, F_{23}\, F_{56} + F_{18}\, F_{27}\, F_{56} +
           F_{38}\, F_{47}\, F_{56} + 
          F_{16}\, F_{25}\, F_{34} + F_{18}\, F_{27}\, F_{34} +\cr
         &  F_{58}\, F_{67}\, F_{34} + 
          F_{36}\, F_{45}\, F_{12} + F_{38}\, F_{47}\, F_{12} -
           F_{58}\, F_{67}\, F_{12} - \cr
             & F_{12} - F_{34} -  F_{56} - F_{78}=0  . \cr }}  
Thus, we have seen that a supersymmetric $p$-cycle in $\IC^p$ can
be characterized as an object which locally satisfies  
the familiar angle  relation $\sum \Phi_j=\pi$.
Solutions to the resulting conditions  for 3-cycles and 4-cycles are not known
so far, but  might  have some impact on our understanding of 
non-perturbative aspects of ${\cal N}=1$ gauge theories along the lines
of \refs{\wittenb,\miemiec}.
It would also be interesting to find the general form of the supersymmetry
conditions for the gauge field strength.

\vskip 1cm

\centerline{{\bf Acknowledgements}}\pano
This work  is supported in part by the European Comission
RTN programme HPRN-CT-2000-00131. We would like to thank A. Uranga
for very helpful discussion.

\vskip 1cm

\listrefs

\bye

\end